\documentclass[aps, prb, twocolumn, showpacs, floatfix,10pt]{revtex4-2}
\usepackage{amsmath}
\usepackage{amsfonts}
\usepackage{amsbsy}
\usepackage{amssymb}
\usepackage{graphicx}
\usepackage{textcomp}
\usepackage[caption=false,singlelinecheck=false]{subfig}
\usepackage{xcolor}
\usepackage{mathrsfs}
\usepackage{mathtools}
\usepackage{bm}
\usepackage{gensymb}
\usepackage{braket,wasysym}
\usepackage[colorlinks,linkcolor=blue,citecolor=red,filecolor=magenta,urlcolor=red,breaklinks]{hyperref}
\usepackage[bottom]{footmisc}
\usepackage{verbatim}
\DeclareMathOperator{\sech}{sech}
\hypersetup{colorlinks=true, urlcolor=blue, citecolor=magenta, pdfborder={0 0 0}}
\usepackage{breakurl}
\usepackage{natbib}

\allowdisplaybreaks

\normalfont
\def\be{\begin{equation}}
	\def\ee{\end{equation}}
\def\bea{\begin{eqnarray}}
	\def\eea{\end{eqnarray}}

\begin{document}
\title{Fractalized magnon transport on the quasicrystal with enhanced stability}

\author{Junmo Jeon}
\email{junmo1996@kaist.ac.kr}
\affiliation{Korea Advanced Institute of Science and  Technology, Daejeon 34141, South Korea}
\author{Se Kwon Kim}
\email{sekwonkim@kaist.ac.kr}
\affiliation{Korea Advanced Institute of Science and  Technology, Daejeon 34141, South Korea}
\author{SungBin Lee}
\email{sungbin@kaist.ac.kr}
\affiliation{Korea Advanced Institute of Science and  Technology, Daejeon 34141, South Korea}

\date{\today}
\begin{abstract}
Magnonics has been receiving significant attention in magnetism and spintronics because of its premise for devices using spin current carried by magnons, quanta of spin-wave excitations of a macroscopically ordered magnetic media. Although magnonics has clear energy-wise advantage over conventional electronics due to the absence of the Joule heating, the inherent magnon-magnon interactions give rise to finite lifetime of the magnons which has been hampering the efficient realizations of magnonic devices. To promote magnonics, it is imperative to identify the delocalized magnon modes that are minimally affected by magnon-magnon interactions and thus possess a long lifetime and use them to achieve efficient magnon transport. Here, we suggest that quasicrystals may offer the solution to this problem via the critical magnon modes that are neither extended nor localized. We find that the critical magnon exhibits fractal characteristics that are absent in conventional magnon modes in regular solids such as a unique power-law scaling and a self-similar distribution of distances showing perfect magnon transmission. Moreover, the critical magnons have longer lifetimes compared to the extended ones in a periodic system, by suppressing the magnon-magnon interaction decay rate. Such enhancement of the magnon stability originates from the presence of the quasi-periodicity and intermediate localization behavior of the critical magnons. Thus, we offer the utility of quasicrystals and their critical spin wave functions in magnonics as unique fractal transport characteristics and enhanced stability.
\end{abstract}
\maketitle

\section{Introduction}
\label{sec: intro}
Magnonics is an emerging field in magnetism that concerns various applications of magnons, the quanta of spin-wave excitations of ordered magnets, in storing, transporting, and processing information\cite{demokritov_2013,prabhakar2009spin,rezende2020fundamentals,chumak2015magnon}.
%In magnonics, the magnon carries the information similar to the electrical charge in the traditional electronics.
Unlike the traditional electronics where the information is carried by electric charges and thus the corresponding information transport generally gives rise to the finite energy dissipation through the Joule heating, magnonics are free from such Ohmic dissipation since the information is carried by magnons that are electrically neutral\cite{schneider2008realization,HIROHATA2020166711}. Because of this practical benefit, magnon transport has been gaining extensive interests from researchers in magnetism\cite{PhysRevLett.112.227201,demokritov2006bose}. Despite such advantages, however, it is still remained as a challenge to improve their lifetime and diffusion length scale\cite{yuan2018experimental,cornelissen2015long}.  
Typically, magnons have a finite lifetime of about ns to $\mu$s, which is mainly governed by magnon-magnon interaction, magnon-phonon interaction, disorder effect, and, in itinerant magnets, magnon-electron interaction\cite{PhysRevB.99.184442,PhysRevB.89.184413,PhysRevB.14.4059,PhysRevLett.109.097201}. A long lifetime is generally required to achieve long-distance spin transport, high quality factor of magnetic resonators and also to integrate magnons in quantum information technology\cite{PhysRevX.3.041023,chen2016spin}. Hence, enhancing a lifetime of magnons is a crucial step to advance magnonics with potential applications\cite{chen2016spin}. 
%Thus, separate from spin transport, there is another fundamental question about the stability of the magnon modes. 
%\tcg{Although the magnon transport characteristics are well studied in ordinary periodic systems obeying unit length scale, the influence of the absence of such unique periodic length scale of the system on the magnon transport has been barely understood. Especially, the quasicrystals which are ordered but not periodic would presents anomalous magnon transport characteristics due to their exotic fractal structure.}

To enhance a lifetime of magnons, it is demanding to establish new magnon characteristics distinct from conventional ones. Here, as a strategy, we suggest a unique solid-state device made of magnetic quasicrystals and discuss their magnonic behavior. 
Unlike the ordinary periodic systems having a unit length scale, quasicrystals refer to ordered but not periodic systems\cite{PhysRevLett.53.1951,suck2013quasicrystals}. Due to the absence of periodic length scale, the magnon exhibits distinct wave function resulting in anomalous transports\cite{roche1997electronic,mayou1993evidence,xing2022quantum,bellissard2000anomalous,jeon2021topological,jeon2022quantum}. 
More specifically, in a conventional periodic system, there are two different types of the wave functions; extended and localized states. First, the localized wave function would not decay in the presence of perturbative interaction. This is because it exponentially decays in space, and hence the opportunity to interact with other states becomes negligible\cite{demokritov_2013,prabhakar2009spin,rezende2020fundamentals,anderson2018basic}. However, the localized wave function cannot be used in transport\cite{cohen2016fundamentals}. 
On the other hand, the extended states contribute to the transport but they could widely experience the decay process by magnon-magnon interactions. Hence, there is the dilemma of choosing an extended magnon or localized magnon for efficient spin transport. To circumvent this dilemma in conventional magnetic crystals, we suggest magnetic quasicrystals and their unique quantum states originated from the exotic quasiperiodic ordering, which are known as the critical states, as alternative spin carriers with enhanced lifetime.
%The quasicrystals which are ordered but do not have any periodic length scale are famous as the Fibonacci quasicrystal in one-dimension. 
Importantly, unlike the conventional crystals, the quasicrystals can admit the unique eigenstate so called a  critical state described as neither extended nor localized but power-law decaying with nontrivial fractal structure\cite{kohmoto1987critical,mace2017critical}. Thus, one may expect that such a critical magnon state can be used to transport with enhanced lifetime compared to the ordinary extended magnons.

In this paper, we provide the unique fractal spin wave transport characteristics and the lifetime enhancement under the quasi-periodically arranged magnetic media. We employ the Fibonacci quasicrystal as a magnetic media whose quasi-periodicity is  encoded in the ratio of two different ferromagnetic XY exchange interactions, which we denote by $J_A$ and $J_B$\cite{liu2016quantum,damanik2016anomalous,hida2005renormalization}. We analytically obtain the spin-wave transmittance in the quasicrystal and clarify their nontrivial fractal characteristics. It turns out that the critical spin transport exhibits a power-law decaying behavior, whose power could be controllable in terms of the strength of quasi-periodicity. Surprisingly, the perfect spin wave transmittance occurs at certain set of length scales, which by themselves exhibit self-similarity. Furthermore, we show that the decay rates of the critical magnons are largely suppressed compared to the case of the extended magnons in the periodic limit and this results in enhancement of magnon lifetime. Our work shows that the critical magnons of the quasicrystals could be a candidate of stable magnetic information carrier in magnonics.

%\section{Quasicrystalline magnet}
%\label{sec: model}
\section{Fractal spin wave and anomalous transports}
\label{sec: transmittance}
Let us consider the ferromagnetic $XY$ spin chain of a quasicrystal under a strong magnetic field along $z$ direction, $h$. The Hamiltonian is given by, 
\begin{align}
\label{XY model}
&\mathcal{H}=\frac{1}{2}\sum_{i=1}^{N-1}J_{i,i+1}(S_i^+S_{i+1}^-+S_{i+1}^+S_{i}^-)+h\sum_{i=1}^{N}S_i^z,
\end{align}
where $J_{i,i+1}$, the exchange interaction between the spins at $i$-th site and ($i+1$)-th site, is arranged quasi-periodically\cite{liu2016quantum,damanik2016anomalous,hida2005renormalization}.
% For the Fibonacci quasi-periodic tiling, the nearest neighbor exchange interaction is given as either $J_A$ and $J_B$.  with $J_A$ and $J_B$. 
The total number of sites is $N$ and $S_i^{+/-}$ is the spin raising/lowering operator at $i$-th site. 
%Within the linearized spin wave theory, we discuss the unique spin wave transport characteristics in the quasicrystalline system.

%\begin{comment}
%The simple linearized spin wave theory originates from the Holstein-Primakoff (HP) transformation which maps the spin operators on the lattice to the bosonic operators $\hat{b}_i,\hat{b}_i^{\dagger}$. With a strong magnetic field, spins are aligned along $z$ direction and the magnon excitation is represented as,
%\begin{align}
%\label{HP}
%&S_i^+=\sqrt{2S-n_i}\hat{b}_i, \\
%&S_i^-=\hat{b}_i^{\dagger}\sqrt{2S-n_i}, \nonumber \\
%&S_i^z=S-n_i. \nonumber
%\end{align}
%Here, $n_i=\hat{b}_i^\dagger\hat{b}_i$ is the magnon number operator. Note that the HP transformation inherently has higher-order terms due to the binominal expansion of $\sqrt{2S-n_i}$ and this leads to the magnon decay. Under the linear spin wave theory, $\sqrt{2S-n_i}$ is approximated into $\sqrt{2S}$, assuming low temperature $k_BT\ll J_{i,i+1}$ and hence $\braket{n_i}\ll S$. Then, the Hamiltonian simply becomes,
%\begin{align}
%\label{linearmodel}
%&H=\frac{S}{2}\sum_{i=1}^{N-1}(\hat{b}_i^\dagger J_{i,i+1}\hat{b}_{i+1}+h.c.)+h\sum_{i=1}^N(S-n_i).
%\end{align}
%With this HP transformation, we first discuss the fractal magnon transport and then consider the reduced magnon decay rate originated from higher-order magnon-magnon interaction terms.  
%\end{comment}

Before discussing the anomalous spin wave transports in a ferromagnetic quasicrystalline magnet, we first characterize the spectrum within the linearized spin wave theory.  Note that the Holstein-Primakoff (HP) transformations map the spin operators to the bosonic operators $\hat{b}_i,\hat{b}_i^{\dagger}$ as $S_i^+=\sqrt{2S-n_i}\hat{b}_i, S_i^-=\hat{b}_i^{\dagger}\sqrt{2S-n_i}, S_i^z=S-\hat{n}_i$ with the magnon number operator $n_i=\hat{b}_i^\dagger\hat{b}_i$\cite{holstein1940field}. 
After the HP transformation, the leading term of Eq.\eqref{XY model} is equivalent to the tight-binding model of bosons with uniform potential energy, $h(NS-1)$ per site that has a sublattice symmetry. Thus, the magnon dispersion should be symmetric in energy with respect to the uniform potential energy\cite{cohen2016fundamentals}. 
%In other words, if the energy $E=h(NS-1)+\varepsilon$ is in the spectrum of the linearized spin wave Hamiltonian, then $E'=h(NS-1)-\varepsilon$ is also in the spectrum.
Throughout the paper, we consider odd $N$ cases, where the energy $h(NS-1)$ is exactly in the middle of the spectrum, thus we term it as the middle energy. 

The middle energy magnon mode is exactly solvable as follows~\cite{mace2017critical}. For the magnon mode with the energy given by $\hbar (N S - 1)$, the Schr\"{o}dinger equation is written as $J_{n+1,n+2}\psi(n+2)+J_{n,n+1}\psi(n)=0$, where $\psi(x)=\braket{\Omega|\hat{b}_x|\psi}$ represents the middle energy magnon mode $\ket{\psi}$ and $\ket{\Omega}$ is the magnon vacuum. Thus, the middle magnon state is generally written as,
\begin{align}
\label{middlestate general}
&\psi(2n+1)=\psi(1)\prod_{k=1}^{n}\left(-\frac{J_{2k-1,2k}}{J_{2k,2k+1}}\right).
\end{align}
For a periodic limit where $J_{i,i+1}=J$, the state is uniform, and hence represents an extended state. If two exchange coefficients $J_A$ and $J_B$ are periodically alternating such as `ABABAB' manner, then the product in Eq.\eqref{middlestate general} blows up or collapses as $n$ increases for $J_A>J_B$ or $J_A<J_B$, respectively. Thus, the middle energy magnon is localized at either $i=1$ or $i=N$, respectively. Note that such localization mode is equivalent to the zero-energy mode of the Su-Shrieffer-Heeger model\cite{su1979solitons}.

On the other hand, in quasicrystals where $J_A$ and $J_B$ are quasi-periodically arranged, the spatial distribution of the middle energy magnon mode exhibits a nontrivial fractal structure\cite{kohmoto1987critical,mace2017critical}. In detail, let $\kappa=\log\rho$ with $\rho=J_A/J_B$ be the strength of the quasi-periodicity. For the pattern-dependent function $a$ defined by $a(\mbox{AB})=1,a(\mbox{BA})=-1$ and $a(\mbox{AA})=a(\mbox{BB})=0$ for the local patterns $AB,BA,AA$ and $BB$, the wave function on the $(2n+1)$-th site is given by
\begin{align}
\label{critical middle}
&\psi(2n+1)=(-1)^n\psi(1)e^{\kappa H(n)}.
\end{align}
Here, $H(n)$ is known as the height field given by $H(n)=\sum_{i=1}^{2n-1}a(w_{[2i-1,2i+1]})$ where $w_{[x,y]}$ is the local pattern from the site $x$ to the site $y$\cite{mace2017critical}. $H(n)$ is pattern-dependent, thus, the middle state given by Eq.\eqref{critical middle} is also pattern-dependent whose dependency is controlled by the strength of the quasi-periodicity, $\kappa$. Moreover, the transmittance is invariant under the exchange of the $A,B$ links which would be encoded by $\kappa\to-\kappa$ and $H(n)\to-H(n)$. Thus, the spin wave function and the characteristics of magnon transport in quasicrystal is solely determined by variation of the height field. 

\begin{figure}[]
\centering
\includegraphics[width=0.5\textwidth]{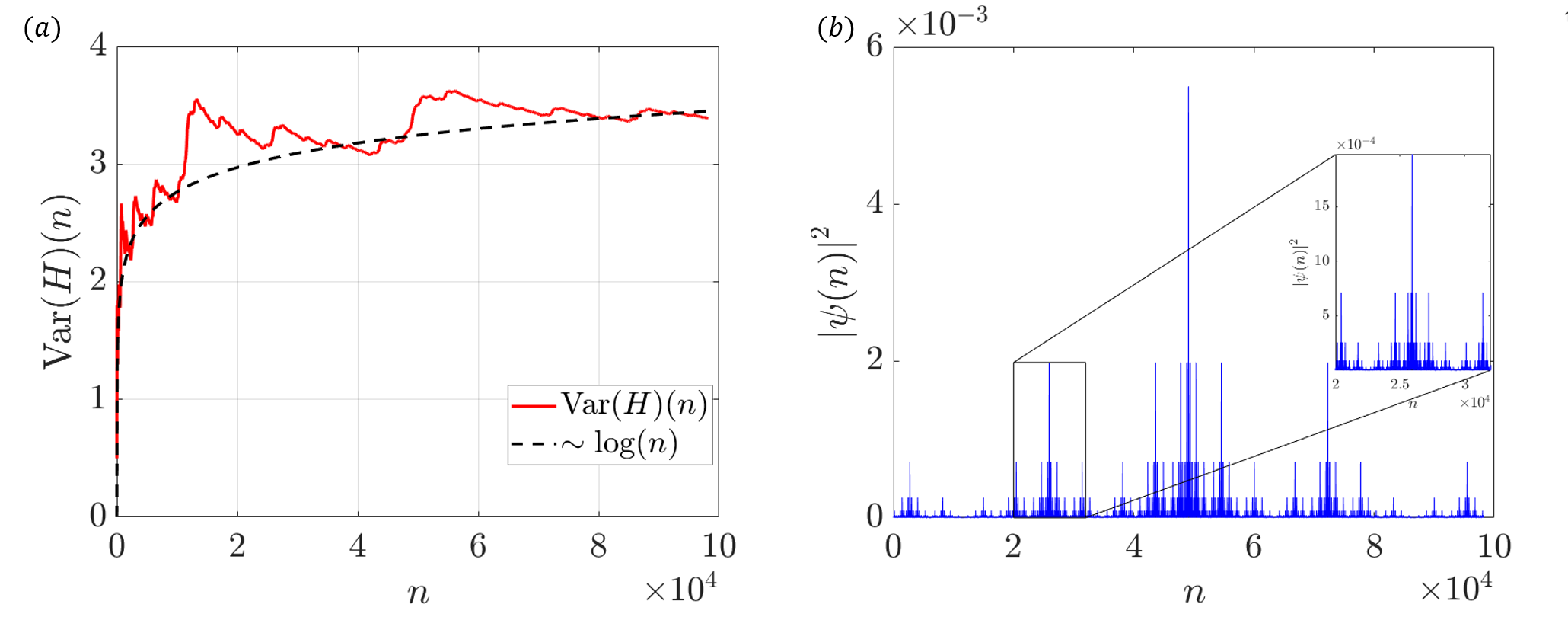}
\label{fig1}
\caption{\label{fig:hfield} (a) Logarithmic scaling behavior of variation of the height field, $\mbox{Var}(H)(n)$, where the argument $n$ indicates the $(2n+1)$-th site. (b) The spatial profile of the middle energy state [Eq.~\eqref{middlestate general}] which is the critical state having the power-law scaling and the self-similar structures. The inset shows the self-similar structure of the spatial profile by zooming in $20000<n<32000$ region. Here, $\rho=J_A/J_B=0.6$ and the system size $N=196419$.}
\end{figure}

For a concrete example, let us consider the Fibonacci quasicrystal where the quasi-periodic pattern resides in two distinct links $A, B$. In particular, the Fibonacci quasicrystal is generated by the successive substitution maps $A\to AB$ and $B\to A$, thus, gives rises to $A\to AB\to ABA\to ABAAB$, and so on\cite{suck2013quasicrystals}. The height field oscillates around zero in the Fibonacci quasicrystal (See Appendix.), and hence its mean value vanishes\cite{jeon2021topological}. However, the variation of the height field would be non-trivial due to the quasi-periodicity of the Fibonacci tiling pattern. Fig.\ref{fig:hfield} (a) illustrates the variation of the height field, $\mbox{Var}(H)(n)=\sum_{i\le n}\left\vert H(i)\right\vert^2/n$. This shows that the variation of the height field $\mbox{Var}(H)(n)$ has the scaling behavior $\mbox{Var}(H)(n)\sim\log(n)$\cite{mace2017critical,jeon2021topological}. Such logarithmic scaling behavior with self-similarity gives rise to the middle energy critical state as shown in Fig.\ref{fig:hfield} (b).

\subsection{Anomalous scaling of magnon transmittance}
\begin{figure}[]
\centering
\includegraphics[width=0.5\textwidth]{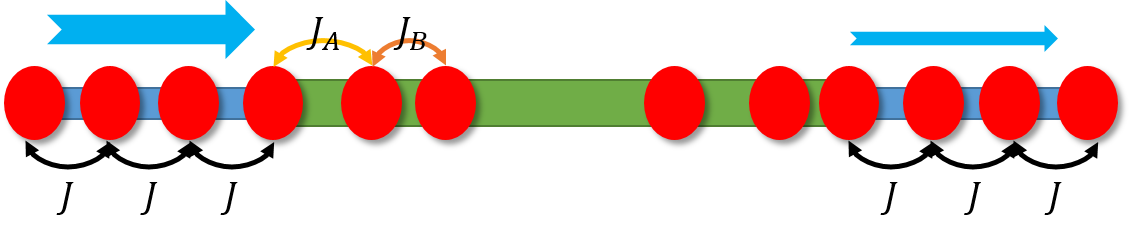}
\label{fig2}
\caption{\label{fig:setup}The schematic experimental setup for the spin wave transport. The two blue regions represent identical semi-infinite periodic leads with uniform exchange magnitude $J$. The green region is the quasicrystal having quasi-periodically arranged exchanges $J_A$ and $J_B$. Each red circle represents the identical magnetic atom. The direction and the width of the blue arrows on the blue regions represent the direction of spin wave transports and the amplitudes of spin waves at specific regions, respectively. The plane spin wave is generated on the left infinity at the specific energy mode.}
\end{figure}

To discuss the unique spin wave transport characteristics in magnetic quasicrystals, let us consider the setup where the quasicrystalline magnet is placed between two semi-infinite periodic spin chains as shown in Fig.\ref{fig:setup}. We generate spin wave propagating from the left to the right as represented by the blue arrows in Fig.\ref{fig:setup}. As passing through the quasicrystal region (green region in Fig.\ref{fig:setup}), the amplitude of the spin wave would be changed.

From the Pichard formula, the transmittance is given in terms of the eigenvalues of the transfer matrix\cite{PhysRevB.40.5276}. In particular, if we generate the middle-energy magnon plane wave from the left side, then the transmittance at the $(2n+1)$-th site is given by\cite{mace2017critical,jeon2021topological}, 
\begin{align}
\label{transmittance}
&T_{2n+1}=\sech^2{(\kappa H(n))}.
\end{align}
If $H(n)$ is a linear function, then the transmittance is exponentially decaying. For zero height field in periodic system, the transmittance becomes uniformly perfect. However, non-trivial height field in quasicrystals leads to the unique characteristics of the magnon transmittance.

First of all, since the typical $H(n)$ behaves as $\sqrt{\log(n)}$ in the Fibonacci quasicrystal, the magnon transmittance decays much slower than the exponential decaying. One can estimate the local dominant power, $\alpha(n)\equiv-\log(T_n)/\log(n)$ of the transmittance. It indicates that the power-law decaying of the transmittance at site $n$. By using Eq.\eqref{transmittance}, we note that $\alpha(n)$ is the decreasing function bounded by $0\le \alpha(n)\le\kappa^2$. Hence, the most dominant decaying power $\alpha(n)$ is spatially dependent and upper bounded by the strength of the quasi-periodicity. Specifically, for stronger quasi-periodicity, the decaying power becomes larger and the transmittance decreases faster. Thus, the strength of quasi-periodicity controls the decay-rate power-law exponent of the transmittance.

%\begin{comment}
%Similar power-law decaying spin transport in the quasicrystal is also emergent in the harmonic mean transmittance, $\braket{T}(n)=n/\sum_{i<n}T_i^{-1}$. It is known that the special distribution of the height field illustrated in Fig.\ref{fig:hfield} (a) gives rise to the power-law decaying $\braket{T}(n)$ as
%\begin{align}
%\label{harmonic mean}
%&\braket{T}(n)\sim \frac{2}{1+n^{F(\kappa)}}.
%\end{align}
%Here, $F(\kappa)$ is an increasing function of the $\kappa^2$, strength of the quasi-periodicity, and $F(0)=0$ since the transmittance in Eq.\eqref{transmittance} is invariant under the exchange of the $A,B$ links. \tcg{(See Supplementary Materials for detailed information of $F(\kappa)$.)} Thus, for given system size, $n$, the harmonic mean magnon transport is power-law decaying in the Fibonacci quasicrystal whose decaying power is controlled by the strength of the quasi-periodicity.
%\end{comment}

\subsection{Self-similar spin wave transport signals}
\begin{figure}[]
\centering
\includegraphics[width=0.5\textwidth]{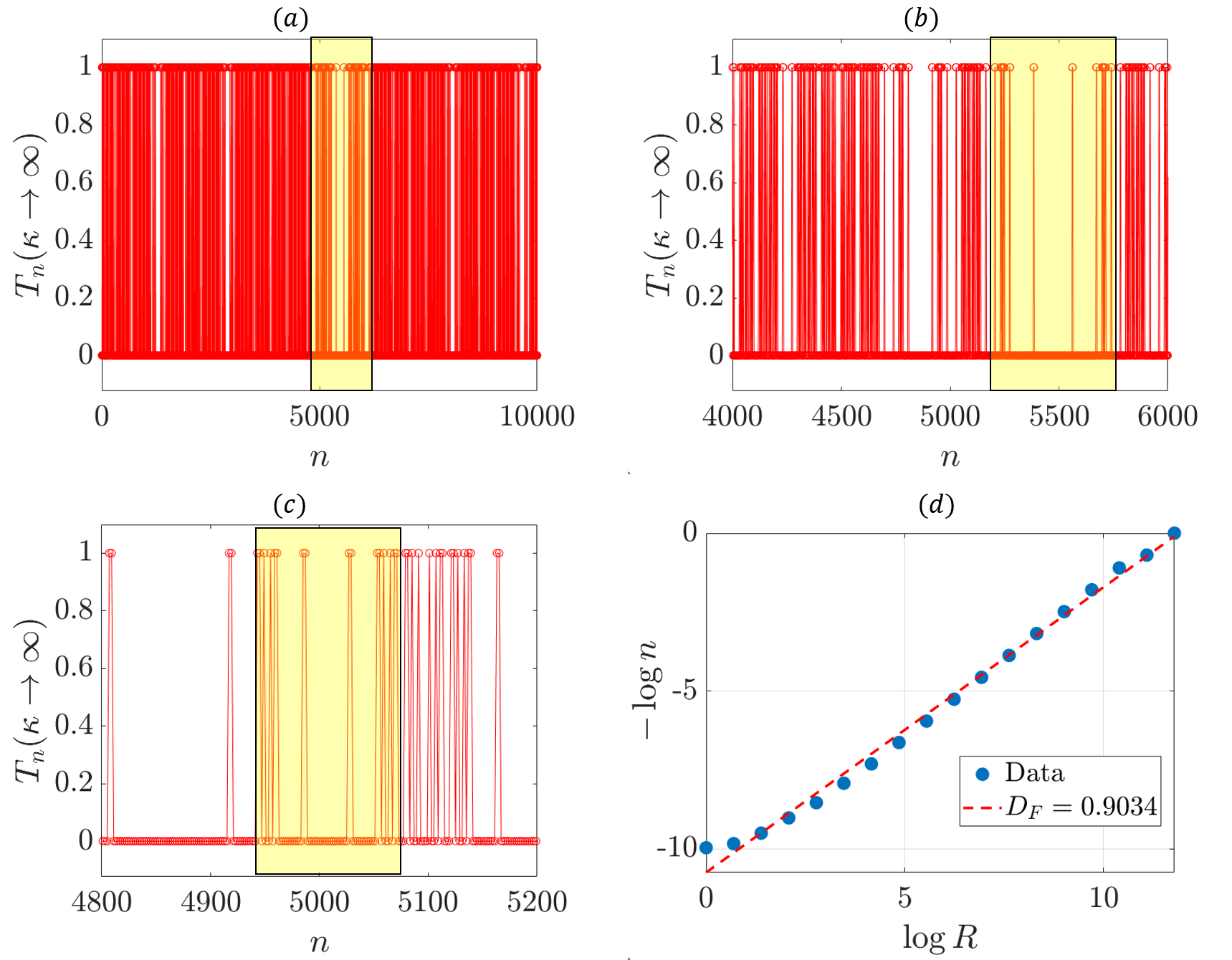}
\label{fig2}
\caption{\label{fig: Fabry}Self-similar distribution of the sites having the perfect transmittance [Eq.\eqref{transmittance}] in the Fibonacci quasicrystal in the limit of the strong quasi-periodicity $\kappa \rightarrow \infty$, where the transmittance at a specific site has the value either 0 or 1. Each panel shows (a) from 0 to 10000, (b) from 4000 to 6000 and (c) from 4800 to 5200 by zooming in five times. The yellow shaded regions exhibit the self-similar pattern of the distribution of the perfect transmittance. (d) Nontrivial Hausdorff dimension, $D_F=0.9034$ indicates the presence of the fractality of the distribution of the sites having the perfect transmittance in the Fibonacci quasicrystal.}
\end{figure}

Now let us consider the case of the strong quasi-periodicity limit given by $\kappa\to\infty$. From Eq.\eqref{transmittance}, the transmittance vanishes except in the case of vanishing height field, $H(n)=0$. Hence, the spin wave signals appear only for these special set of the positions having a zero height field. The number of sites having a perfect transmittance grows as the system size increases (See Appendix.).
%, thus, there are infinitely many sites that admit the perfect transmittance in thermodynamic limit. 
Nevertheless, their distribution would exhibit a nontrivial fractality as the self-similar distribution of the height field.
%For the Fibonacci quasicrystal, the growth of the number of the sites having the perfect transmittance is linear whose slope is $0.2152$.
Fig.\ref{fig: Fabry} shows the transmittance distributions for strong quasi-periodic limit. Interestingly, the positions showing the perfect transmittance for any $\kappa$ form a self-similar structure themselves. Specifically, the yellow regions in Figs.\ref{fig: Fabry} exhibit a self-similar structure. 
%Thus, in the quasicrystals, the self-similar spin wave transport signals appear for strong quasi-periodic system as the perfect transmittances.

To quantify the fractality in the distribution of the perfect transmittance, we compute its Hausdorff dimension. Let $n$ be the number of intervals whose length $R$ needed to cover the region where perfect transmittance appears. Then, the Hausdorff dimension is given by $D_F=-\frac{\partial(\log{n})}{\partial(\log{R})}$\cite{lapidus2004fractal}. Absence of the fractality gives the trivial Hausdorff dimensions, 0 or 1. On the other hand, the distribution of the prefect transmittance gives nontrivial Hausdorff dimension, $D_F\approx 0.9034$. Fig.\ref{fig: Fabry} (d) illustrates the the nontrivial Hausdorff dimension of the distribution of sites having a perfect transmittance.

\section{Decay rate suppression in quasicrystals}
For magnon applications, the stability of magnons with a long lifetime is one of the most important issues\cite{demokritov_2013,prabhakar2009spin,rezende2020fundamentals,chen2016spin}.  
Generally, the magnon modes have a finite lifetime originated from various decaying process such as magnon-magnon or magnon-phonon interactions\cite{PhysRevB.99.184442,PhysRevB.89.184413,PhysRevB.14.4059,PhysRevLett.109.097201}. Particularly, we focus on how the magnon-magnon interaction affects to the stability of the critical magnon mode in the Fibonacci quasicrystal.
% Because the critical magnon is neither localized nor extended, one could expect that the decay rate is suppressed for the critical magnons compared to the extended modes. 
To explore the multi-magnon interaction,  the higher-order terms of HP transformations are considered\cite{holstein1940field,rezende2020fundamentals}. At the fourth order, the HP transformation of Eq.\eqref{XY model} gives rise to, 
\begin{align}
\label{HP4}
&H_4=S\sum_{i=1}^{N-1}J_{i,i+1}\left(\frac{1}{4}\hat{b}_i^{\dagger}(n_i+n_{i+1})\hat{b}_{i+1}+h.c.\right).
\end{align}
%where the total Hamiltonian $\mathcal{H}=H+H_4$ where $H$ is the linearized Hamiltonian in Eq.\eqref{XY model}.
The Hamiltonian commutes with the total magnon number operator, $\hat{\mathcal{N}}=\sum_{i=1}^{N}n_i$ i.e. $[\mathcal{H},\hat{\mathcal{N}}]=0$. Therefore, $H_4$ leads to the two magnon interaction preserving the number of magnons in the ferromagnetic system.

From the Fermi's golden rule, the decay rate $\Gamma$ originated from $H_4$ is given by,
\begin{align}
\label{Gamma}
&\Gamma_{i\to\{f\}}=\frac{2\pi}{\hbar}\sum_{f}\left\vert \braket{f|H_4|i} \right\vert^2 \rho_2(E_f).
\end{align}
Here, $\rho_2(E_f)$ is the two-particle density of states (DOS) for the total energy $E_f$. $\ket{i}$ and $\ket{f}$ are the initial and the final states and their energies are the same $E_f= E_i$. %Since $H_4$ is time-independent, $E_f=E_i$ i.e. energy should be conserved.

Because of the energy conservation during the decay, if there is no $\ket{f}\neq \ket{i}$ state such that $E_f=E_i$, the decay rate in Eq.\eqref{Gamma} vanishes. To avoid such trivial cases, we consider the case that the number of possible final states is maximized. Since the linearized bare Hamiltonian has the sub-lattice symmetry, there are $(N+1)/2$ many particle-hole pairs of the magnon modes whose energies are $E_{\pm}(\varepsilon)=h(NS-1)\pm\varepsilon$, and hence the total energy is the middle energy. Thus, in general, if we consider the two-particle sector of the bosonic Fock space, the middle energy subspace has the maximal degeneracy with respect to the linearized bare Hamiltonian.

Let us consider the decay of two middle energy magnon modes. The initial and final states are given by $\ket{i}=\frac{1}{\sqrt{2}}(\hat{b}_{E_0}^\dagger)^2\ket{\Omega}$ and $\ket{f}=\hat{b}_{E_{+}}^\dagger\hat{b}_{E_-}^\dagger\ket{0}$ unless $E_\pm\neq E_0$, middle energy. Here, $\hat{b}^\dagger_\varepsilon$ is the bosonic creation operator for the energy eigenstate with energy $\varepsilon$ of the bare Hamiltonian. To capture the effect of the quasi-periodicity rather than the mere magnification of the interaction strength, we keep the total magnitude of the exchange interaction strength which appears directly in the decaying rate in Eq.\eqref{Gamma}. In detail, we keep $\sum_{i=1}^{N-1}J_{i,i+1}$ constant with $N-1$ the number of the links so that the average exchange interaction strength is $J$. When the numbers of $A$ and $B$ types of links are $N_A$ and $N_B$, respectively, the proper $J_A$ and $J_B$ are $J_B=J(\rho N_A+N_B)^{-1}$ and $J_A=\rho J_B$ for a given strength of the quasi-periodicity, $\rho$\cite{jeon2022pattern}.

On the other hand, the two particle DOS at the total energy window $[E,E+dE]$ is given by $\rho_2(E)=\int_{E_{min}}^{E_{max}} \rho_1(\varepsilon)\rho_1(E-\varepsilon)d\varepsilon$, where $\rho_1(\varepsilon)$ is the one particle DOS at the energy window $[\varepsilon,\varepsilon+d\varepsilon]$. In our case, $E=2E_0$ where $E_0$ is the middle energy. For one particle DOS, we use the local density of states, LDOS which is given by $\mbox{LDOS}(i,\varepsilon)=-\sum_{n}\mbox{Im}\left( \frac{|V(i,n)|^2}{\varepsilon_n-\varepsilon+i0^+} \right)$. Here, the unitary matrix $V$ is given by $H=VDV^\dagger$ with the diagonal matrix $D$, $V(i,n)$ is the $(i,n)$-element of the unitary matrix $V$, and $\varepsilon_n$ is the energy of $n$-th energy level of $H$. Then, the total single particle DOS for the energy window $[\varepsilon,\varepsilon+d\varepsilon]$ is given by the spatial trace of LDOS.

\begin{figure}[]
\centering
\includegraphics[width=0.4\textwidth]{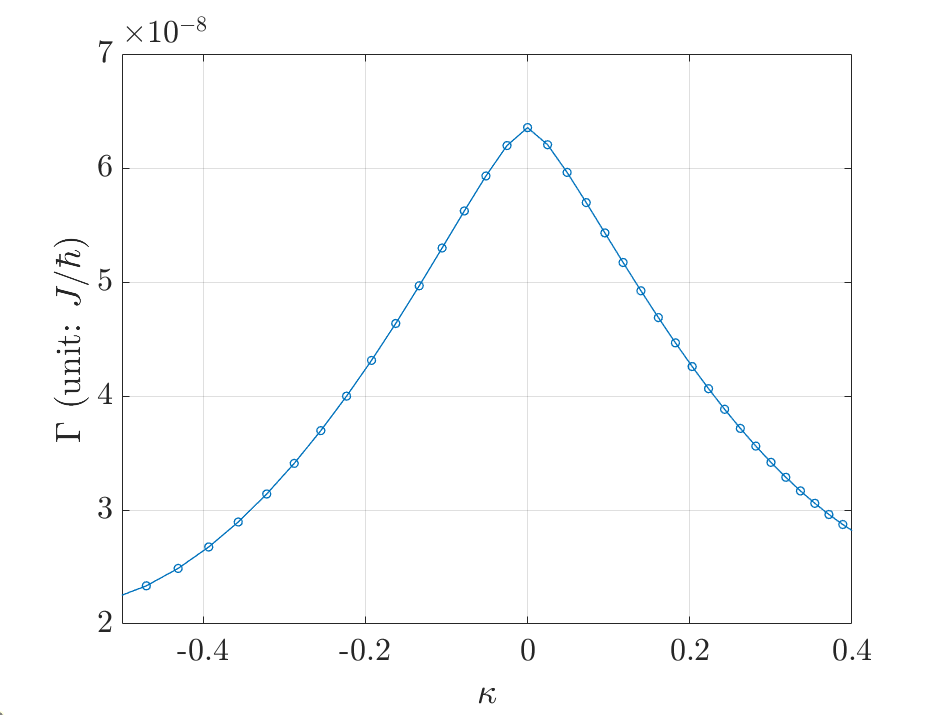}
\label{fig1}
\caption{\label{fig:decayrate} Decay rate, $\Gamma$ as the function of the strength of the quasi-periodicity, $\kappa$. The strength of the quasi-periodicity increases as deviating from the periodic limit, $\kappa=0$. As increasing the strength of the quasi-periodicity, the decay rate is suppressed. The initial and final states in the considered decaying processes are given by $\ket{i}=\frac{1}{\sqrt{2}}(\hat{b}_{E_0}^\dagger)^2\ket{\Omega}$ and $\ket{f}=\hat{b}_{E_{+}}^\dagger\hat{b}_{E_-}^\dagger\ket{0}$ with $E_\pm\neq E_0$ where $E_0$ is the middle energy. The considered system size is $N=10947$.}
\end{figure}

Fig.\ref{fig:decayrate} shows how the decay rate changes in the presence of quasi-periodicity. 
As a function of quasi-periodicity $\kappa$, the decay rate $\Gamma$ is suppressed. 
Here, the system size $N=10947$ for each $\rho=J_A/J_B$. The decay rate, $\Gamma$ is maximized for the uniformly periodic limit where $J_A=J_B$. For the periodic limit, the initial states are both extended states, while for all other cases that $\rho\neq 1$, the initial states are both critical states. Since the critical states have the intermediate spatial distribution neither extended nor localized, they also admit the intermediate decay rate between the extended and localized states. Thus, the magnon decay rate of the critical state becomes smaller than the case of the extended state.

Note that the two particle DOS, $\rho_2(E)$, also contributes to the decay rate. Since the quasi-periodicity makes the energy spectrum flatten and create many gaps known as a fractal spectrum, the DOS would increase in quasicrystals\cite{kohmoto1987critical}. Nevertheless, the decay rate is suppressed. Hence, we conclude that the magnons at the middle energy have a longer lifetime due to their fractal spatial distribution as the critical states.

%\tcb{With the enhanced lifetime in the quasicrystal, the critical magnons can have the high Q-factor FMR in the magnonics. Note that the the Q-factor of a resonator indicates that the relative loss of the energy compared to the stored one in the resonator. Higher Q-factor means smaller energy loss compared to the stored energy in the resonator. Thus, the high Q-factor FMR is one of the important parts of the magnonics and stable spin engineerings. Since the Q-factor is proportional to the lifetime of the resonator, the longer lifetime of the magnon mode tells us more stable energy storing in the magnonics.}

\

\section{Discussion and Conclusion}
\label{sec:con}
In conclusion, we have established the self-similar spin wave transport with the enhanced stability in quasicrystals. Focusing on the critical magnon state at the middle of the spectrum, we have demonstrated that the spin wave transport is neither uniform nor exponentially decaying. Instead, it shows the power-law decaying behavior. In addition, we have shown that the magnetic quasicrystal can admit the perfect transmittance at special set of lengths. The distribution of such sites for perfect transmittance is independent on the strength of the quasi-periodicity, and exhibits a self-similar structure with the nontrivial Hausdorff dimension. Furthermore, we have shown that the critical magnons are more stable under the magnon-magnon interaction compared to the conventional periodic case, resulting in a longer lifetime. Our findings suggest that the critical magnon states in magnetic quasicrystals may serve as long-lifetime spin carriers that are required to advance magnonics. The effects of other factors on magnon lifetime such as magnon-phonon interaction and impurities in quasicrystals would be an interesting future work. 

Our work shows that magnetic quasicrystals can offer magnons with unusual characteristics that cannot be found in traditional magnetic crystals. It leads us to speculate that they might provide useful functionalities as well in current-driven spintronics, exhibiting, e.g., certain special form of spin-transfer torque or spin-orbit torque that are hard to achieve with periodic magnetic systems\cite{chen2016spin,pallecchi2006current,manchon2009theory,vzelezny2018spin}. More generally, we believe that magnetic quasicrystals would enrich the material library of magnonics and spintronics  and thereby facilitate the advancement of the both fields.

\subsection*{Acknowledgement}
J.M.J. and S.B.L. are supported by National Research Foundation Grant (No. 2020R1A4A3079707, No. 2021R1A2C1093060),). S.K.K. were supported by Brain Pool Plus Program through the National Research Foundation of Korea funded by the Ministry of Science and ICT (NRF-2020H1D3A2A03099291).

\bibliography{my1}

\clearpage
\newpage 

\appendix
\renewcommand\thefigure{\thesection\arabic{figure}}
\setcounter{figure}{0}

\begin{widetext}
\section{Oscillating behavior of the heigth field in the Fibonacci quasicrystal}
\label{subsec:supple1}
In this appendix, we briefly show that the average value of the height field in the Fibonacci quasicrystal vanishes. The Fibonacci quasicrystal is comprised of two letters $A,B$ with the substitution maps $A\to AB, B\to A$. These substitution maps can be written in a matrix form $S=\begin{pmatrix}1&1 \\ 1&0\end{pmatrix}$. Since $BB$ is forbidden under these substitution maps, there are only three types of the length-2 supertiles in the Fibonacci quasicrystal, explicitly, $AA, AB$ and $BA$. Let us denote them as $X,Y$ and $Z$ respectively. Now, let us renormalize the Fibonacci quasicrystal with these length-2 supertiles. For example, under this renormalization procedure, the first few part of the Fibonacci quasicrystal, $ABAABABA$ becomes $YXZZ$. Note that the pattern-dependent function, $a$ is given by $a(X)=0, a(Y)=1$ and $a(Z)=-1$. Thus, the height field at the site $n$, $H(n)$ the cumulative sum of the function $a$ up to the site $n$ is given by $N_Y-N_Z$. Here, $N_Y, N_Z$ are the number of the $Y$ and $Z$ length-2 supertiles up to the site $n$.

We claim that $N_Y-N_Z$ vanishes on average in the thermodynamic limit. To show this, we obtain the substitution maps for $X,Y,Z$ supertiles induced by the original substitution maps for $A,B$ tiles. By applying the original subsitution maps in three times for $X,Y,Z$ supertiles, respectively, one can easily get the new subsitution maps, $X\to YXZZY$, $Y\to YXZZ$ and $Z\to YXZY$. These substitution maps for the supertiles would be given by the matrix $S'$ in Eq.\eqref{seq1}
\begin{align}
\label{seq1}
&S'=\begin{pmatrix}1&1&1 \\ 2&1&2 \\ 2&2&1\end{pmatrix}.
\end{align}
Here, the basis is given by $N_X,N_Y,N_Z$, the numbers of the $X,Y,Z$ supertiles in the tiling. The eigenvalues of the matrix $S'$ are $-1$ and $2\pm\sqrt{5}$. Especially, the eigenvalue $-1$ corresponds to the eigenvector $(0,1,-1)^T$. This eigenvector corresponds to the $N_Y-N_Z$. The eigenvalue $-1$ indicates that the quantity $N_Y-N_Z$ is oscillating keeping its magnitude under the substitution map. Thus, in thermodynamic limit, the mean value of the height field would be negligible on average as desired.

One of the interesting point is that there are infinitely many $n$ has the zero-valued height field. Because the height field is oscillating around the zero on average, but the pattern-dependent function $a$ increase or decrease the height field by 1. Thus, at least one site between the oscillation of the height field takes zero value. Hence, in the thermodynamic limit, we have infinitely many sites of the sites having the zero-valued height field which admit the perfect transmittance regardless of the strength of the quasi-periodicity. See the main text for the unique self-similar distribution of these sites in the Fibonacci quasicrystal.
\end{widetext}

\end{document}